# A Survey of Requirement Engineering Process in Android Application Development

Ali Nawaz
Department of Computer & Software Engineering,
CEME, NUST
Islamabad, Pakistan
alinawaz7587@gmail.com

Attique Ur Rehman
Department of Computer & Software Engineering,
CEME, NUST
Islamabad, Pakistan
ahattique@gmail.com

Wasi Haider Butt
Department of Computer & Software Engineering,
CEME, NUST
Islamabad, Pakistan
wasi@ceme.nust.edu.pk

## ABSTRACT
Mobile application development is the most rapidly growing industry in the world. Nowadays, people totally depend on smart phones for performing daily routine tasks which results in tremendous rises in the expectation of human being from IT industry which increase the requirements of human being. In order to tackle the uncontrolled changes in the requirements, IT experts performed some proper requirement engineering process (REP). Therefore, in this paper we are performing industry survey by asking them several questions related to the REP from android developer in order to understand the REP used in the IT industry. Results we extract from this study is satisfactory used in order to make REP more effective.

## 1. INTRODUCTION
Mobile technology revolutionized the daily life of human beings by making everything digital and available in their hands.  Today we can pay our utility bills, make booking, access whole world via social media and anything we wish by just using our fingers. From last two decades, there is tremendous increasing in the number of mobile devices companies, mobile models and operating system as well. According to GSMA [1], there is 7,740,979,987 world population and 5,156,156,519 are active mobile subscribers which means approximately 66% of the world population using mobile phone. Such a huge number of users attract many stakeholder and market thinkers to invest their time, efforts and resources in order to achieve business goals. When we look at the mobile industry there are numerous companies available like Samsung, iPhone, Huawei, Oppo and each company owes millions of users. Such a huge amount of user and devices also attract device makers to create operating system like android, iOS, blackberry, windows. According to reports, Android is most widely used operating system with 65% of overall user which is followed by iOS with 32% user then followed by window phone and so on.There are several programming environment and tools are available for mobile application development. Android application development is usually performed in android studio [2] which is official Integrated Development Environment (IDE) support by Google. Since 2019, Kotlin for android application development is preferred over Java and C++. For iOS [3], Swift is used which is the powerful, intuitive, precise and expressive programing language used for iOS development. For window phone, developer use Microsoft Visual Studio environment. Similarly, there are some IDEs for Blackberry, Symbian and Java ME. In addition to this there are some cross-platform [4] application development like Microsoft's Xamarin, Flutter, Adobe PhoneGap, Corona, Appcelator and many more.

It is observed that mobile industry is the most rapidly growing and dynamic IT industry [5] and software is the backbone of IT industry. It is stated that "small program can be written without using software engineering but without using software engineering principles it would be difficult to develop large programs". Therefore, it is clear need to follow software engineering models for android application development it will not only reduce time and cost but also helps to tackle uncontrolled changed in requirements and improve quality. As we know that the first and foremost step of software development lifecycle is Requirement analysis or REP and all the next activities of life cycle relying on it that is why we need to handle requirements efficiently. REP is composed of two sub processes [6] i.e. Requirement development and Requirement management. Requirement development is composed of four activities. Elicitation is the process to gather the requirements from stakeholder by using elicitation techniques. Analysis is the process to understand the information gather from elicitation. Specification is the process to check whether the necessary requirements are included or not. Validation is the last step of requirement development and performed to check whether all requirement defined by stakeholders are valid or not. Requirement management is the other sub process of the REP and concerned with the management of requirements.

A survey is conduct which cover the REP used in android application development industry of Pakistan, Australia and USA. This survey will cover many aspects of requirement development activities used in REP. This survey will be beneficial for both Requirement Engineer and Android application developer.

## 2. LITERATURE REVIEW

In [7], conduct an industry survey in which they select different small and medium size organization and asked few open questions regarding the requirement engineering. They conclude with the fact that every organization need to develop their own REP, need to improve their existing RE processes and perform automation of current requirement engineering practices. The survey is not suitable for current IT industry as it was conducted in year 2000.

In [8] conduct an industry survey among Malaysian requirement engineering practitioners. The survey technique used by them are questionnaires and mailing. The survey reveals that requirement engineers still facing organizational and technical challenge in order to get requirements. The survey also suggests the use of REP improvement framework called R-CMM [9].

Also [10] tries to identity the state of art approach used for REP in Chile. They use questioner's method to understand the REP and their issues. The survey identified that in addition to ad-hoc RE practices used in the RE process, lacking of communication between client and customer is the major hurdle in the REP. The conducted survey is limited to small organization of Chile.

Similarly, in [11] conduct an industry survey comprises of 15 software companies of Pakistan. The main techniques used by them are administrative questionnaires and interview. The survey concludes that requirement process in software companies of Pakistan is vary from each other with respect to business and organization level. The survey also reveals that poor requirement process lead to poor results which in some cases results in redesigning and redevelopment of software. The survey suggest that requirement engineer should pay his full attention and work hard in order to perform requirement process successfully and which in turn results in the successful development of software product.

[12] conduct a survey regarding the requirement engineering used in small and medium size software houses of Pakistan and method used by them in order to conduct a survey is questionnaires. The purpose of their survey is to show the level of requirement engineering principles followed by software houses in Pakistan. They conclude that small and medium size software houses don't follow any standard of requirement engineering and requirement processes used by them is immature and need clear enhancement. The number of survey respondents are 17.

## 3. METHODOLGY

The literature related to the REP in mobile application development is quite less and relatively few works has been done so far. So, it is decided to conduct industry survey of Android Applications development in different IT firms and freelancer of Pakistan, Australia and USA. The questionnaire method is used to get responses from the individuals and companies. We send google form, conduct interview and send hard copy to the respondent in order to perform this survey. Instead of asking many irrelevant questions from respondent we directly asked the questions related to the topic and receive quite well response from the respondents. We specially requested the respondent to not fill the questionnaires individually but should take advice from industry experts or do some peer-review. After receiving the response author did some peer review with individual respondent to remove the complexity in the response form. Following are questions we raises in the survey:

RQ 1: Which Requirement Elicitation technique you use?

RQ 2: How much effort spent on Requirement analysis with respect to whole REP?

RQ 3: Effect of REP on project resources?

RQ 4: Percentage of conflicts resolved during Requirement analysis?

RQ 5: Which Criteria do you use to validate Requirements?

In first question we asked the respondent about his background. Now in modern development, the developers are categorized into two broad categories whether they are employee of some organization or freelancers So, we asked them about their employment status.

Second question is also related to the respondent background in which we asked them about their experience is android development. This question particularly reflects the accuracy of response made by respondent. This is multiple choice question and the choices are categorized into four categories.

Third question is general question related to REP in which we asked about the using of REP for android development. The is also nominal question and the available choices are Yes, No and sometimes.

In third question we specifically asked that out of interview, meeting, workshops and questionnaires which requirement elicitation technique [13] is usually used by them in order to get the requirement.

The fourth question is related to the second subprocess [14] of REP in which we asked them about the percentage of time spent on analysis with respect to whole requirement engineered process. The choices available to the respondents are ranges from 0% to 80%.

Next question we asked from the respondent is that by applying REP, what will be the effect of it on projects resources whether resources are within time and budget, slipped time and within budget, slipped time and over budget or within time and over budget.

The major goal of requirement analysis is to find conflicts so, in second last question we asked respondent about percentage of conflict resolved during requirement analysis. The percentage of conflicts might range from 0% to 100% so, the options available to the respondent are ranges from 0% to 100%.

In last question, we asked about the validation criteria used by the respondent. There are lots of evaluation criteria for requirement validation [15] but in this survey, we present three option which are requirement conformance with business goal, conformance with SRS and conformance with technical and financial feasibility.

These are major research questions we raise in our research survey. We send this survey to more than 60 individuals and 15 different software development firms. The response receives from individual who might be the employee of some firm is very impressive than response from organization. Out of 75 we received response from approximately 45% individual developer and software organization

## 3.1 Survey Analysis and Results:

### 3.1.1 Respondent Profile

The figure shows the employment status of the android developer. There are two broad categories of employment status that are freelancers and company employers. From last 2 decades Global Software Development (GSD) [16] is widely used for accessing the multiple skills and other resources from across the world by using the several outsourcing platforms like Upwork [17], fiverr [18], freelancer etc. Although there are certain issues like time and cultures are involved in GSD but still GSD is widely used and most successful approach for development software because of major advantage of accessing skill around the world. Most of the people in underdevelopment countries like Pakistan prefer to work in some organization because they don't want to take any risk regarding employment.

Employment Status?

27 Responses

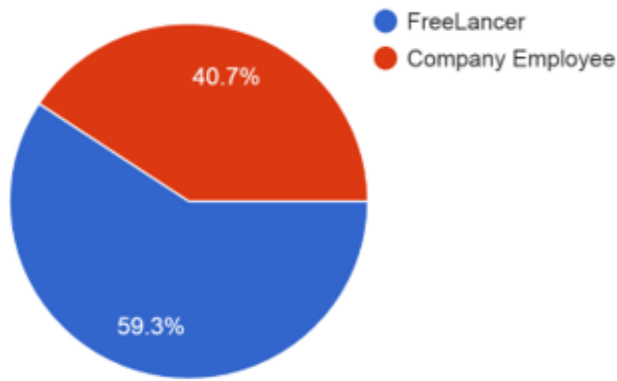

**Figure 1. Employment status**

The 11 respondents out of total respondents are company employees which are 40.7% of total employment status means that they are the representative of software development organization. The 16 participants are freelancers and are global software developer which are 59.3% of total percentage.

### 3.1.2 Development Experience

The first android phone [19] was lanced by HTC in 2008 which means that android application development has not very long history and assumed to be the advance field in modern technology in fact it is responsible for advancement in modern world and technology. In this question we asked participants about their experience in the field and question asked is MCQ format and choices are ranges from fresh to 10 years. The responses received from the participants are quite expected. The responses clearly show that with the launch of new IDEs and new platforms the number of users and developers of particularly software increases. It also reveals the fact that with the launched of android studio [20] in 2013 the number of android developers is raises and increases with the passage of time.

Your Experience?

27 Responses

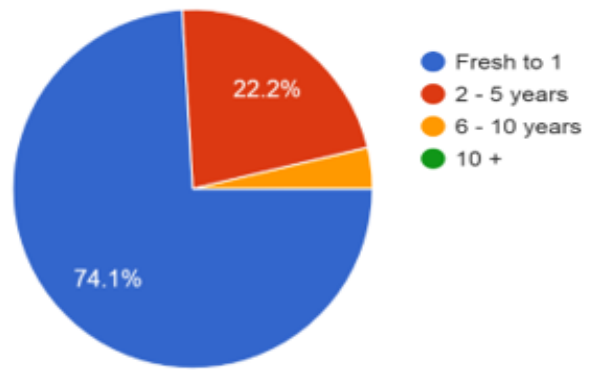

**Figure 2. Respondant Experience**

The responses of the participants clearly reveal the fact that developer of android applications are increases with the passage of time which is directly proportional to the number of users of the android application. It shows that 20 participants have less than 1 years' experience 6 participants have experience of between 2 and 5 years and only one participant has experience of between 6 to 10 years.

### 3.1.3 REP in Development

It is usually said that a small program can be written without using software engineering but program of size more than 1000 lines of codes is difficult to write without using software engineering. Generally, it is observed that complete software is developed without using software engineering it is like collecting requirement on some paper and start development but some professional software developers as well as prominent CMMI and ISO standard organizations follows complete SDLC as well as REP in which they properly elicit, analyse, specify and validate requirements. So, in this question we asked them about the usage of REP in android apps development and received expected responses.

Do you use requirement engineering processes in development?

27 responses

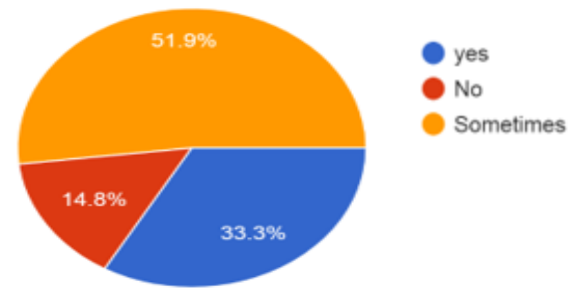

**Figure 3. REP in development**

The responses reveal the fact that without using requirement we can develop android application. 14 participants respondent that they sometimes use REP 9 participants reveals they use REP fluently and 4 participants doesn't use REP in android application development and these are usually freelancers.

### 3.1.4 Elicitation Technique

Elicitation is the first process of requirement engineering and concerned with collecting the need and determining the problem of user. It is assumed to be most critical process because all SDLC is based on it. There are certain elicitation techniques used in software and the most prominent are interview, meeting, workshop and questionnaires each technique has pros and cons. The responses received clearly shows that there is no best elicitation technique in software. In this survey we asked respondent that out of prominent elicitation technique which technique is widely used by them.

Which Requirement Elicitation technique you use?

27 Responses

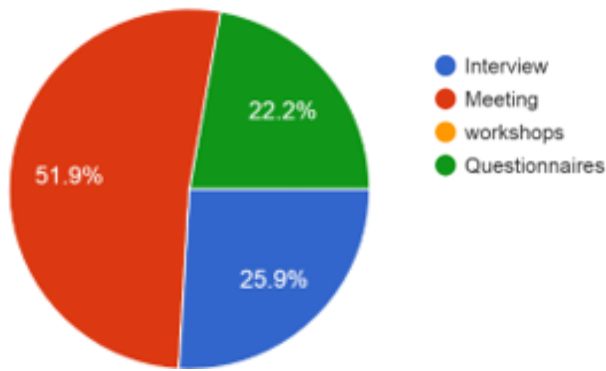

**Figure 4. Elicitation technique**

The responses show that there is no best elicitation technique. It clearly reveals that 14 participants use meeting technique; interview is used by 7 participants and 6 participants use questionnaires.

### 3.1.5 Efforts Spent on Requirement Analysis

Analysis is the second activity of REP and is most time-consuming activity of the process. In this activity the elicit requirements are processed and trying to understand it. Requirement is specifically related to skills of team members and resources required to performed it that is why we simply asked the respondents about the effort spent on analysing the requirement. The responses received are mixed and reveals that it depends on the software being developed.

How much effort spent on Requirement analysis with respect to whole Requirements engineering processes?

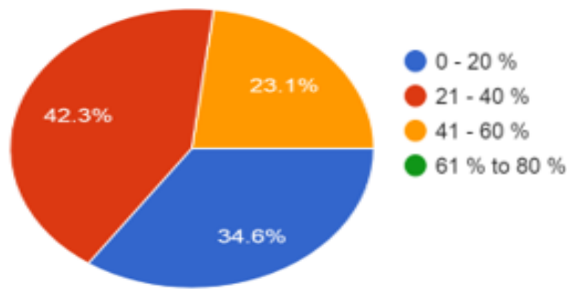

**Figure 5. Effort spent on Requirement Analysis**

This is also MCQ and the choices available to the respondent ranges from 0 to 80%. It reveals that 11 participants spent 21 to 40% of efforts, 9 participants spent the efforts ranges from 0 to 20% and 6 participant spent effort between 44 to 60%. This also conclude that effort spent on elicitation is always than 60%.

### 3.1.6 REP Effects on Requirement Resources

The major goal of software engineering is the developing of quality product within specified time and budget [21]. Time and budget are the important resources of the software. In this question we asked about the effect of REP on project resources. The response we obtained from this question is surprising.

Effect of Requirement Engineering processes on project resources?

27 Responses

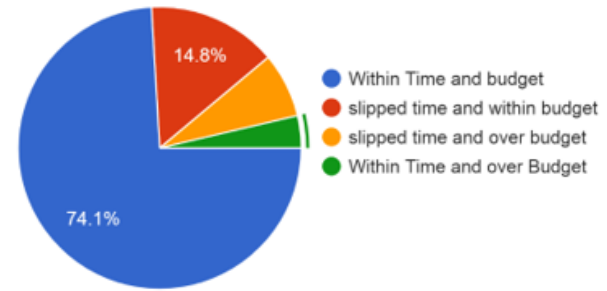

**Figure 6. Effect of REP on project resources**

In this question we asked that by applying REP whether the application developed is Within Time and budget, slipped time but within budgets, slipped time but over budget, within time but over budget. The total of 20 participants reveals that by applying REP the developed software is within time and within budget, 4 participant reveals that software developed is within budget but slipped time, 2 participants reveals slipped time and over budget and only 1 participant choose within time and over budget.

### 3.1.7 Conflict Resolution

Conflicts in the requirements are mostly occur and conflicts are most difficult to find and resolve. The ultimate goal of requirement analysis activity is to resolve requirement conflicts. In this question we asked respondent about the conflict they find and resolved during process. The responses received are quite effective and reflect the importance of applying REP and requirement analysis activity.

Percentage of conflicts resolved during Requirement analysis?

27 Response

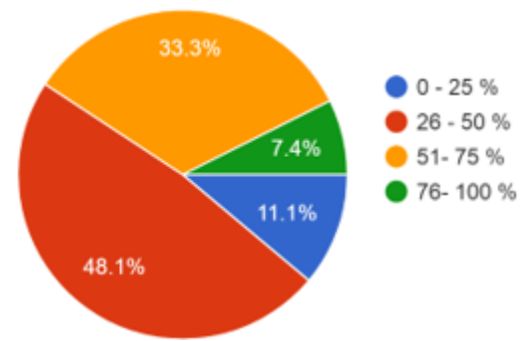

**Figure 7. Total Conflicts resolved**

The choices presented to the respondent ranges from 0 to 100%. The response reveals that 13 participants finds the conflict between 26 to 50%. 9 participants find the conflicts between 51 and 75%. 3 participants find conflicts between 0 to 25% and remaining 2 participants find conflicts between 76 to 100%. The means that requirement analysis is the efficient and effective activity for conflict resolution [22].

### 3.1.8 Validation of Requirements

The last activity of REP is to check the validation of elicited, analysed and specified requirements. In this question we specifically asked about the validation criteria [23] used by android application development. There are certain validation criteria of requirement but in this survey, we limit the response of respondent onto three norms. As all software specifically android applications are developed for user so, the first criteria are to check whether it meets the user needs or not. The second criteria are conformance with business goal and third criteria is conformance with technical and financial feasibility.

Which Criteria do you use to validate Requirements?

27 Responses

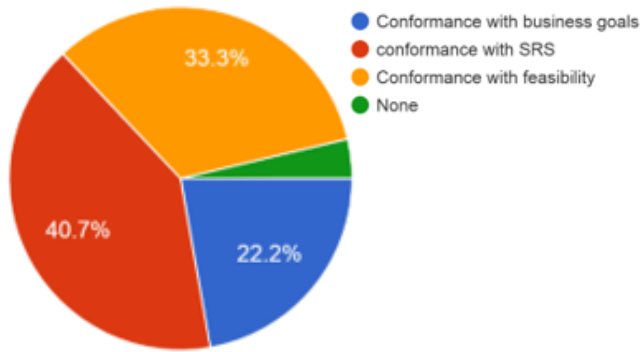

**Figure 8. Requirements Validation criteria**

The 11 participants out of 27 choose the conformance with SRS criteria which is 40.7% of overall percentage. 9 participants choose conformance to feasibility criteria, 6 participants choose conformance with business goal criteria and 1 participant doesn't choose any option.

## 4. DISCUSSION

Quantitative analysis of survey reveals that most of the responders are the freelancers in software development term we called them global software developer. This result is expected because the number of software development by traditional approach is less than global software development. The android development has not very long history and interest in android development increase when google launched android studio and making it publicly available in 2013. This attract developers to enters into new regime of development that is why most of the respondents are fresh in the field of android development and this response also reveals the fact that with the passage of time android developers and enthusiastic are increasing. It is observed from the experience of authors as well as reveal by the survey that SDLC as well as REP is not used by android developers but well reputed organization and professional software developers used well defined SDLC and REP that fact is more clarify in this survey it reveals that sometimes REP is followed in android application development. The first activity of REP is the gathering of requirement and it is the most critical activity because overall SLDC relying on it. This activity is concerned with the way of gathering the requirement so, it is observed that one to one meeting is the most widely requirement elicitation technique followed by interview and questionnaires. After eliciting the requirements, the next activity is to process them to understand them. It is the activity specifically related to the effort of developer so it is revealing that individual android developer or organization generally spent 21 to 40% effort on this activity. This ultimate goal of analysis is to find conflict statements in the requirements and this survey reveals this fact that the percentage of conflicts find and then resolved by this activity is more than 50%. The major objective of software engineering is to developed quality software on time and budget that is why it is assumed that every work performed on the software should utilize the resources. This survey reveals the fact that by following REP the product developed is mostly within budget and time which means that this process doesn't affect the project seriously. Sometimes, it may result in slipping of time and over budget [24] by purely depends on the nature of project. There are many criteria for validating the requirement as REP is concerned only with requirement so, it is reveals in the survey that requirement validation is to check whether requirement is conformance with SRS meanwhile few participants also perform validation to check conformance with feasibility of financial and technical and conformance with business goal because the main goal of software engineering is the developing of quality product within time and budget.

## 5. CONCLUSION AND FUTURE WORKS

In this paper, the survey is conduct regarding REP in android applications development and participants of the survey includes android developers, business analyst, project managers and requirement engineering. Most of the android developers are freelancers having experience ranging from fresh to 1 years. Large organization and professional android developers usually follows requirement engineering as well as SDLC for large software development and ignored by small organization that is why REP is sometimes used in the industry. Requirement elicitation is the basepoint of REP as well as whole SDLC and meeting, interview and questionnaires are the most commonly requirement elicitation technique used for deliberate searching of requirement. Analysis is concerned with finding the conflicts and it is observed that more than 50% conflicts resolved during this activity. Software engineering deals with developing quality product on specified time and other resources and it is also revealed in the survey that REP doesn't affect the time and budget. The whole REP applied to remove conflicts and then validate the final draft of requirements So, it is observed that goal of this process is to find conformance of requirements with SRS which in turn results in the conformance of business goal. Due to limitation of resources, we conduct this survey on small scale involves few participants and obtain very little data. In future, we are interested to gain large data and applying data mining and machine learning techniques to achieve accuracy in classification of different activities of REP.

## 6. ACKNOWLEDMENT

The authors really recognise the effort of respondents who spent time from their busy schedule to fill our survey.

**AUTHORS' BACKGROUND**

| Your Name | Position | Email | Research Field | Personal website |
|---|---|---|---|---|
| Ali Nawaz | Master student | anawaz.cse19ceme@ce.ceme.edu.pk | Software Engineering Machine learning and Data mining | |
| Attique Ur Rehman | Master student | aurehman.cse19ceme@ce.ceme.edu.pk | Software Engineering, Software management and Requirement engineering | |
| Wasi Haider Butt | Assistant Professor | wasi@ceme.nust.edu.pk | Software Engineering and Requirement Engineering | |